\documentclass[runningheads]{llncs}
\usepackage{graphicx}
\usepackage{tabu}
\usepackage{booktabs}
\usepackage{multirow}
\usepackage{array}
\usepackage[utf8]{inputenc}

\usepackage[table]{xcolor}
\usepackage{hhline}
\usepackage{pgf}
\usepackage{tikz}
\usepackage{collcell}
\usepackage{adjustbox}
\usepackage{array}
\usepackage{array}
\usepackage{multirow}

\newcommand\MyBox[2]{
  \fbox{\lower0.5cm
    \vbox to 1.7cm{\vfil
      \hbox to 1.7cm{\hfil\parbox{1.4cm}{#1\\#2}\hfil}
      \vfil}%
  }%
}

\usepackage{xcolor}
\definecolor{accent_pink}{RGB}{248,33,110}
\definecolor{accent_blue}{RGB}{19,149,186}
\definecolor{accent_green}{HTML}{4fa400}
\definecolor{accent_orange}{HTML}{ff8a39}
\definecolor{accent_yellow}{HTML}{f8ca00}
\colorlet{highlight_accent}{accent_yellow!50!white}

\usepackage{todonotes}
\usepackage{soul}
\usepackage{changes}
\setaddedmarkup{\color{accent_green}\emph{#1}}
\setdeletedmarkup{\color{accent_pink}\sout{#1}}

\begin{document}
\title{Estimation of preterm birth markers with U-Net segmentation network}
%
%
\author{Tomasz W\l odarczyk\inst{1}, Szymon P\l otka\inst{1, 5}, Tomasz Trzci\'nski\inst{1, 4}, Przemys\l aw Rokita\inst{1}, Nicole Sochacki-W\'ojcicka\inst{2,3}, Micha\l\space Lipa\inst{2}, Jakub W\'ojcicki\inst{2,3}}

\authorrunning{T. W\l odarczyk et al.}

\institute{Warsaw University of Technology \and Warsaw Medical University \and  Ernest W\'ojcicki Prenatal Medicine Foundation \and Tooploox \and MedApp S.A.}
\maketitle              

\begin{abstract}
Preterm birth is the most common cause of neonatal death. Current diagnostic methods that assess the risk of preterm birth involve the collection of maternal characteristics and transvaginal ultrasound imaging conducted in the first and second trimester of pregnancy. Analysis of the ultrasound data is based on visual inspection of images by gynaecologist, sometimes  supported  by  hand-designed  image  features  such  as  cervical  length. Due to the complexity of this process and its subjective component, approximately  30\%  of  spontaneous  preterm  deliveries  are  not correctly  predicted. Moreover, 10\% of the predicted preterm deliveries are false-positives \cite{C1}. In this paper, we address the problem of predicting spontaneous preterm delivery using machine learning. To achieve this goal, we propose to first use a deep neural network architecture for segmenting prenatal ultrasound images and then automatically extract two biophysical ultrasound markers, cervical length (CL) and anterior cervical angle (ACA), from the resulting images. Our method allows to estimate ultrasound markers without human oversight. Furthermore, we show that CL and ACA markers, when combined, allow us to decrease false-negative ratio from 30\% to 18\%. Finally, contrary to the current approaches to diagnostics methods that rely only on gynaecologist's expertise, our method introduce objectively obtained results.
\end{abstract}
\section{Introduction}
Preterm birth (PTB) affects 5-18\% of pregnancies worldwide, which is equivalent to 15 million preterm neonates each year \cite{C1}.
Despite major advances in perinatal care, preterm birth still accounts for 75\% of neonatal deaths and over 50\% of neurological handicap in children \cite{C2}. Preterm birth is defined as birth before 37 weeks of gestation, however high mortality and morbidity mainly affects neonates delivered before 34 weeks, often referred to as early preterm (1-3\% of all pregnancies) \cite{C3}. Prediction and early detection of women at high risk of PTB are crucial as it allows timely intervention. Despite potentially effective treatments like cervical cerclage, vaginal progesterone or pessaries, accurate, early diagnosis still remains a major challenge \cite{C4}--\cite{C9}.
Current screening methods combine maternal characteristics, obstetric history and cervical length measured at 20-24 weeks \cite{C3}. A major disadvantage of this approach lies in failing to identify women with cervical incompetence before the second trimester and therefore missing the opportunity for successful intervention. Attempts have been made at validating the same screening markers in the first trimester with variable results, the best yielding a detection rate of 54.8\% at a false-positive rate of 10\% \cite{C11}.

In this paper, we address the problem of spontaneous preterm birth prediction.
We present a novel method for estimating two biophysical ultrasound markers: cervical length (CL) and anterior cervical angle (ACA). Cervical length marker refers to the length of the lower end of uterus. Anterior cervical angle is defined by angle between the uterine wall and the cervical canal. We introduce additional feature - ACA marker - for preterm birth prediction as suggested by the results published in \cite{C16}. Extending \cite{C16}, we computed ACA automatically and combined the results with the CL marker, what significantly improved the overall prediction quality. To achieve that goal, we use a deep neural network architecture trained for segmenting prenatal ultrasound images. To overcome the fact that our ultrasound dataset, after balancing procedure, is very small and it could be a vital reason for poor performance, we decide to use a different dataset to perform prediction, to what is described in Sec. 3.3.
Finally, we present that in comparison to regular analysis of ultrasound data, our method performs better and can be used to obtain different biophysical markers as well. 

\begin{figure}[ht!]
    \centering
    \includegraphics[width=12cm]{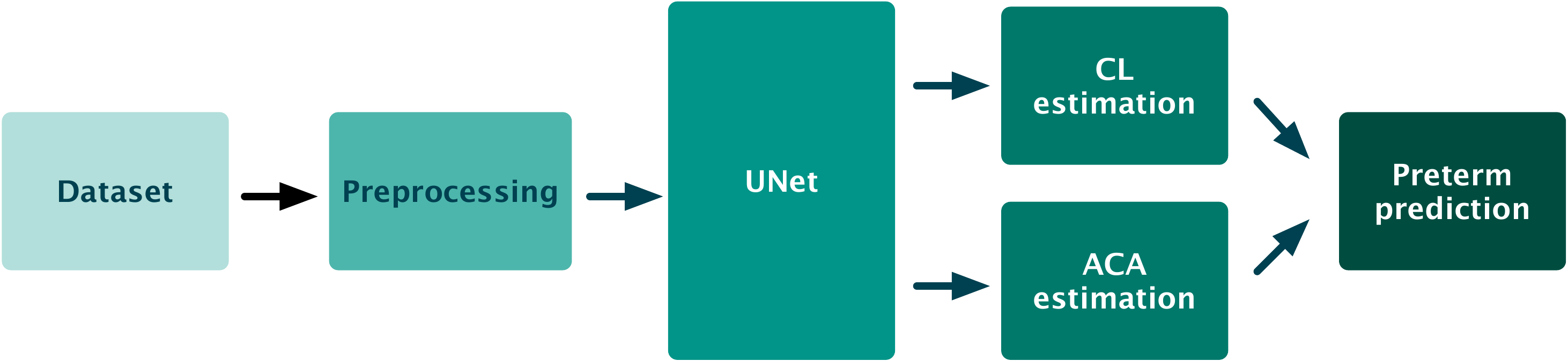}
    \caption{The proposed workflow of estimation preterm birth markers. Our method after data preprocessing uses the U-Net network for segmentation of the cervix, and then allows the estimation of CL and ACA markers.}
    \label{fig:fig-pipeline}
\end{figure}

\section{Method}

\begin{figure}[ht!]
    \centering
    \includegraphics[width=12cm]{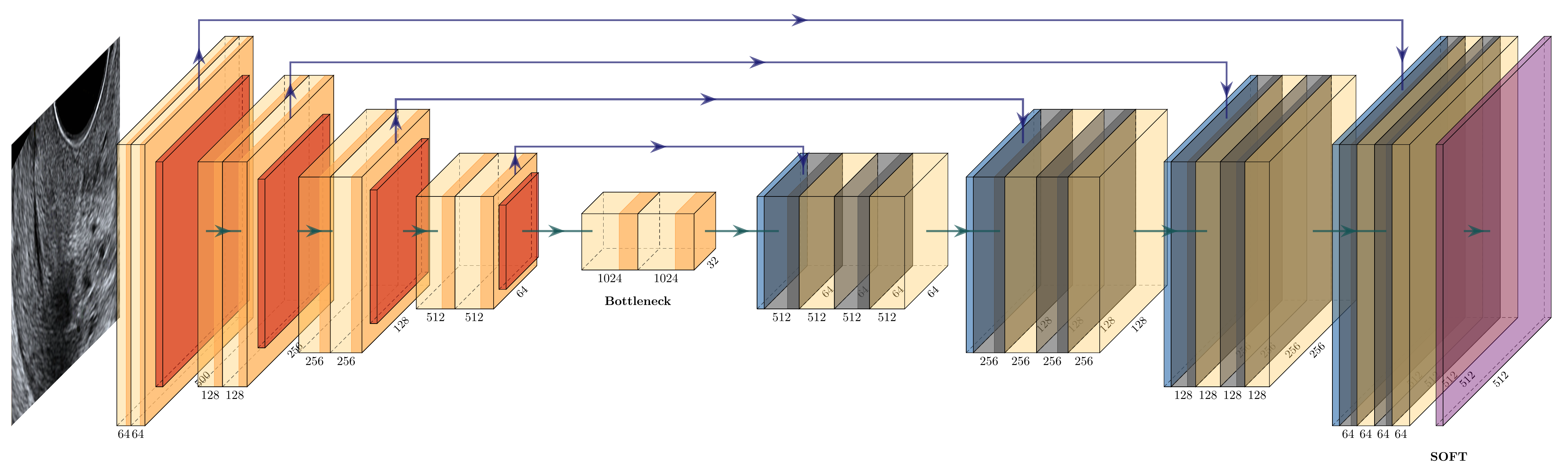}
    \caption{The U-Net architecture \cite{C13}. Each box represents feature maps. The number of channels is signed under each feature map.}
    \label{fig:fig-arch}
\end{figure}

In this section we present our method of estimation of CL and ACA markers that relies on cervix extraction with U-Net segmentation, as depicted in Fig. \ref{fig:fig-pipeline}. The U-Net \cite{C13} architecture is an encoder-decoder neural network implementation used for semantic segmentation, mainly designed for biomedical image processing. This architecture is illustrated in Fig.\ref{fig:fig-arch}

We start training a U-Net model for the segmentation task of extracting a cervical shape from ultrasound images. Once trained, we use our neural network to obtain binary masks of the cervix. Finally, we use them to estimate CL and ACA markers and then for binary classification task (preterm vs. control). To perform cervical length estimation we apply the centerline algorithm \cite{C17} to the binary masks. Such algorithm relies on a generation of a Voronoi diagram for given cervix shape to get the polygon skeleton where the skeleton centerline is selected and smoothed. We use the same extracted masks for ACA estimation with different approach based on a recurential split on centroid location for a given shape. 

\section{Experiments}

In this section, we present results obtained with the proposed method. We first describe the dataset used in our experiments and show the results obtained using the segmentation algorithm. We then verify if the estimated CL  and ACA metrics correspond to the ground truth one. In the second part we evaluate whether CL and ACA combined, perform better than current methods and present results of the classification task (preterm \textit{vs} control).

The first stage in our workflow is cervical segmentation using the U-Net neural network. The segmentation results are used to estimate CL and ACA described in the second stage.

\subsection{U-Net segmentation}

\begin{figure}[ht!]
    \centering
    \includegraphics[width=12cm]{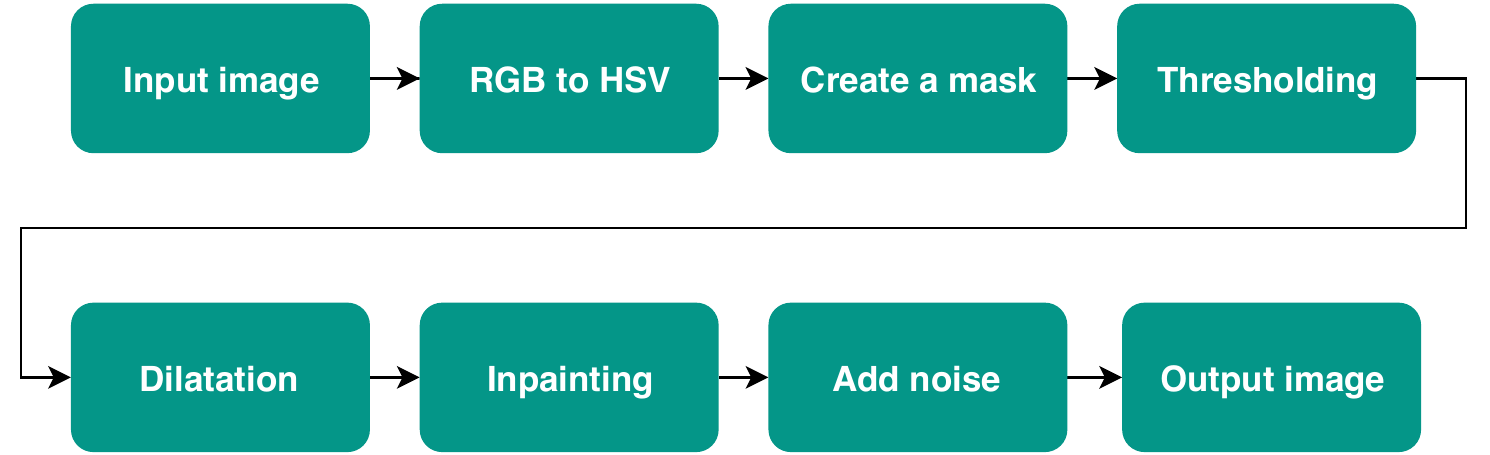}
    \caption{Data preprocessing flow.}
    \label{fig:fig-inpaiting}
\end{figure}

\textbf{Dataset and Preprocessing:} The data collection was collected at King's College London and Warsaw Medical University and it contains data from 359 pregnant women with 316 control pregnancies and 43 preterm deliveries, which is defined as birth before 37 weeks of gestation. The data was registered and labeled using standard infrastructure for ultrasound imagery operated by specialized physicians. Since our dataset contains images (and not the raw data), the annotations are embedded in the graphical layer and hence cannot be filtered automatically out of the data. To overcome this shortcoming and prevent U-Net from focusing only on annotated markers we decide to remove all annotations from images using inpainting method. Inpainting methods using machine learning did not give satisfactory results on our ultrasound images, so we use standard computer vision algorithms. At first we convert our dataset from the RGB to the HSV colour space.
Next, we define the range of colours of all annotations in the HSV space, what allows us to detect these ones which we want to get rid of. The next step is to create a mask. Then through thresholding we obtain a binary image based on defined color range. We then use dilation (a morphological operation on the image) to expand our mask to completely remove annotations around the extracted pixels in the first step.
The inpainting method was used in order to prevent the U-Net network from focusing on coloured markers in the images. The diagram of the method described above is presented in Fig \ref{fig:fig-inpaiting}.

\begin{figure}[ht!]
\begin{minipage}[b]{.48\linewidth}
  \centering
  \centerline{\includegraphics[width=5.5cm]{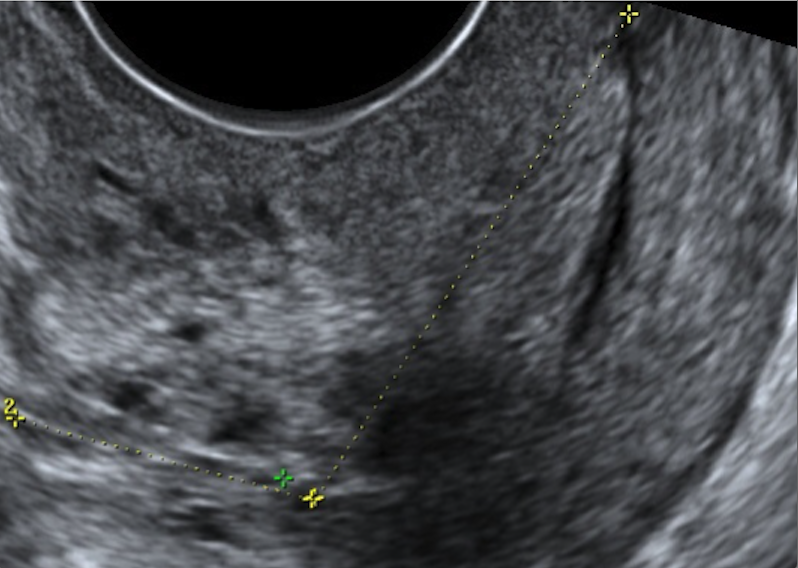}}
  \centerline{(a) Image with annotations}\medskip
\end{minipage}
\hfill
\begin{minipage}[b]{0.48\linewidth}
  \centering
  \centerline{\includegraphics[width=5.5cm]{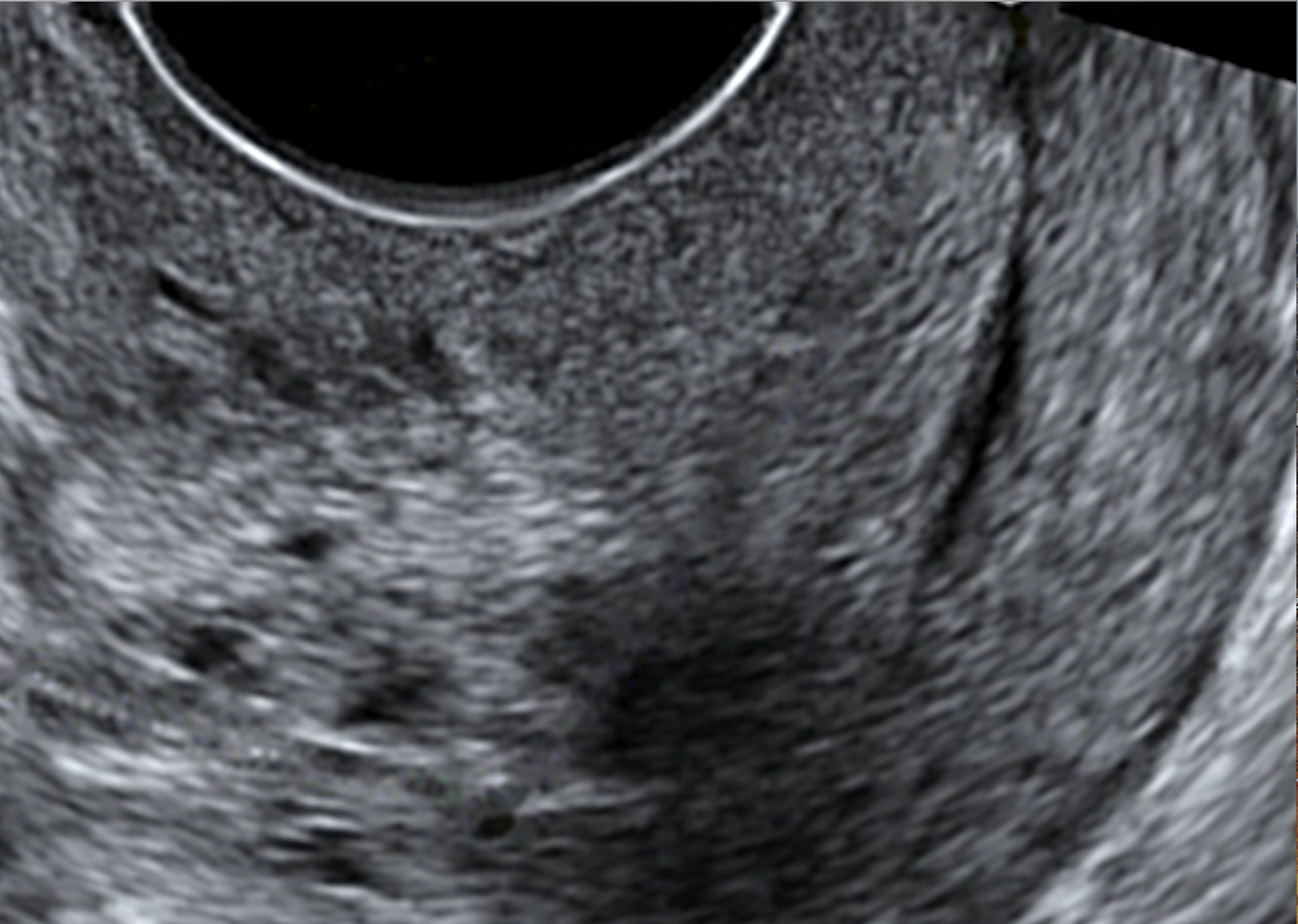}}
  \centerline{(b) Image without annotations}\medskip
\end{minipage}
\caption{Example of using our inpainting method. On the left, we presented the original image, and on the right after applying our inpainting algorithm. Our method was used in order to prevent the U-Net network from focusing on coloured markers in the images.}
\label{fig:res}
\end{figure}

The dataset contains around 20\% preterms which reflects the statistical occurrence of this phenomenon in reality. To mitigate this shortcoming we balanced the dataset by applying data augmentation to achieve a 50:50 ratio, to avoid heavily focusing on the majority class by classification algorithm. We augmented the dataset to 6359 images (359 original and 6000 augmented) by random
rotations in the range of -10 to 10 degrees, random contrast and brightness adjustments. We divide it into training and validation subsets maintaining a ratio of 70:30. \newline

\begin{figure}[ht!]
  \centering
  \centerline{\includegraphics[width=8cm]{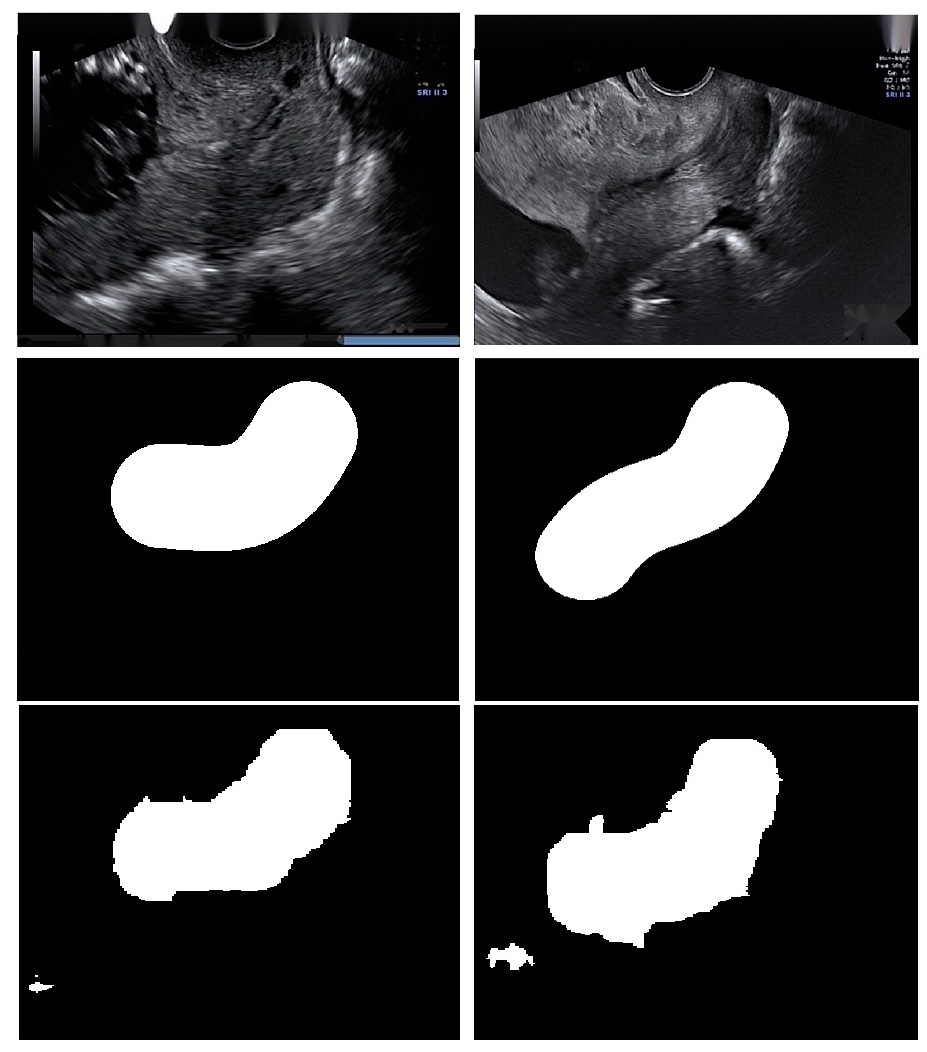}}
\caption{Segmentation results on our dataset. We present from top to bottom: input image after removing the annotation from the original images, ground truth and prediction after applying the U-Net network. Our method allowed us to achieve an average Jaccard index of 0.91, a minimum of 0.89 and a maximum of 0.92, with a standard deviation of 0.1.}
\label{fig:unet_results}
\end{figure}

\noindent\textbf{Experimental settings:} We use our augmented dataset to train a network on a machine with AMD FX-8320 @ 3.5Ghz CPU and NVIDIA TITAN X 12GB GPU. We implement our models using the PyTorch library with CUDA support.
We train U-Net for 650 epochs with a batch size of 4, Adam optimizer with a learning rate of $10^{-4}$ and weight decay of $10^{-4}$. We use BCEWithLogits as a loss function.
We use the 256 px $\times$ 256 px images as input while initializing weights with Xavier uniform method (also known as Glorot initialization) with $\sqrt{2}$ gain.
\newline
\hfill \break
\noindent\textbf{Binary segmentation mask:} We evaluate the U-Net neural network on the task of cervix segmentation of the dataset. We use \textit{Jaccard Index}, also known as \textit{Intersection over Union (IoU)} as the evaluation metric during training. For two sets A and B, the Jaccard index is defined as the following:

\begin{equation}
    J(A,B) \ =\ \frac{|A\cap B|}{|A\cup B|}
\end{equation}
For cervix segmentation task we obtain average Jaccard Index of 0.91 (min - 0.89, max - 0.92, SD - 0.1). Several results are presented in Fig. \ref{fig:unet_results}. In the optimisation of the neural network, we controlled for both Dice and Jaccard index, but more consistent results were obtained with the Jaccard index.

\subsection{CL and ACA estimation}

\begin{figure}[ht!]
\begin{minipage}[b]{.48\linewidth}
  \centering
  \centerline{\includegraphics[width=7.0cm]{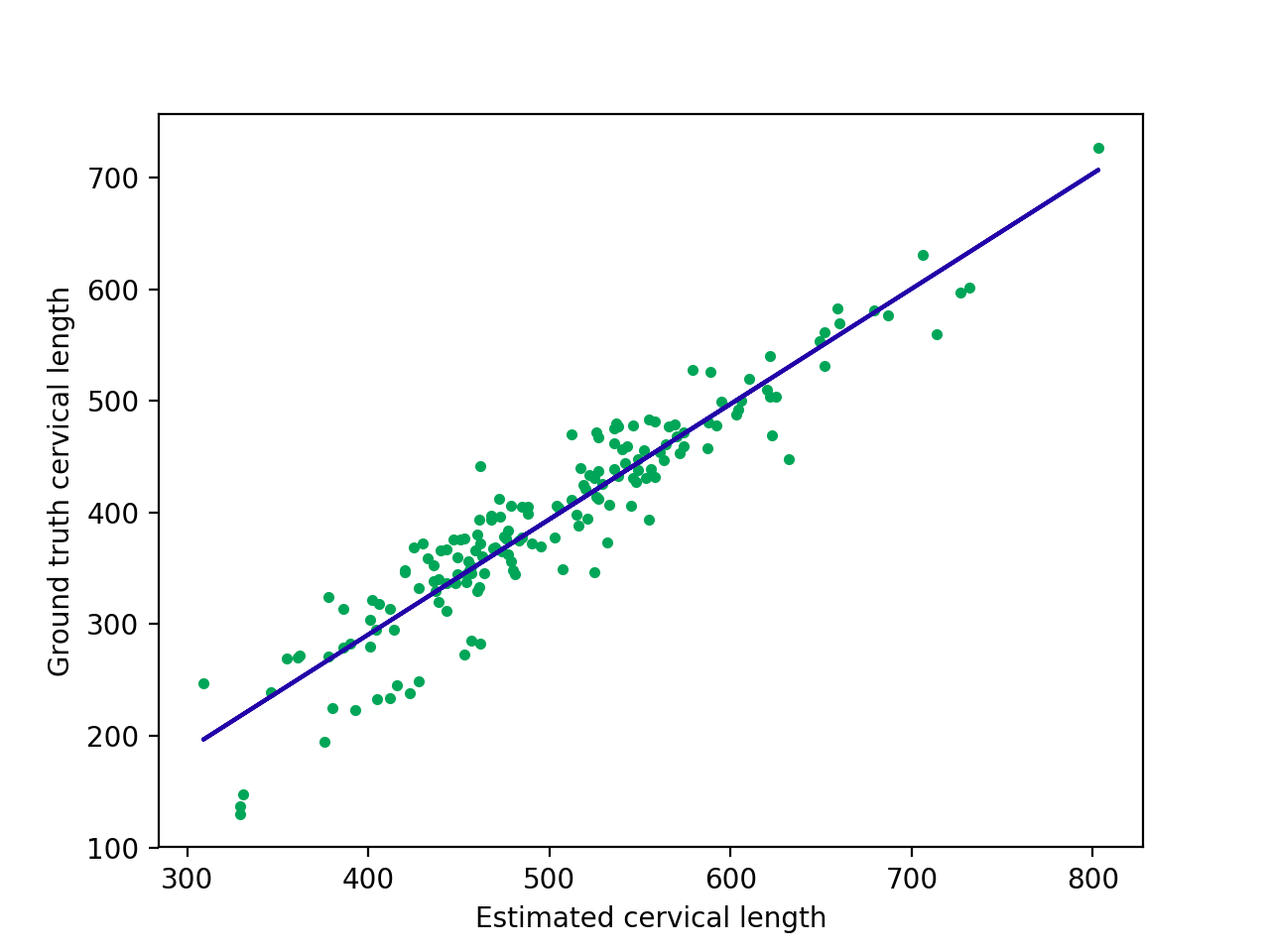}}
  \centerline{(a) CL estimation}\medskip
\end{minipage}
\hfill
\begin{minipage}[b]{0.48\linewidth}
  \centering
  \centerline{\includegraphics[width=7.0cm]{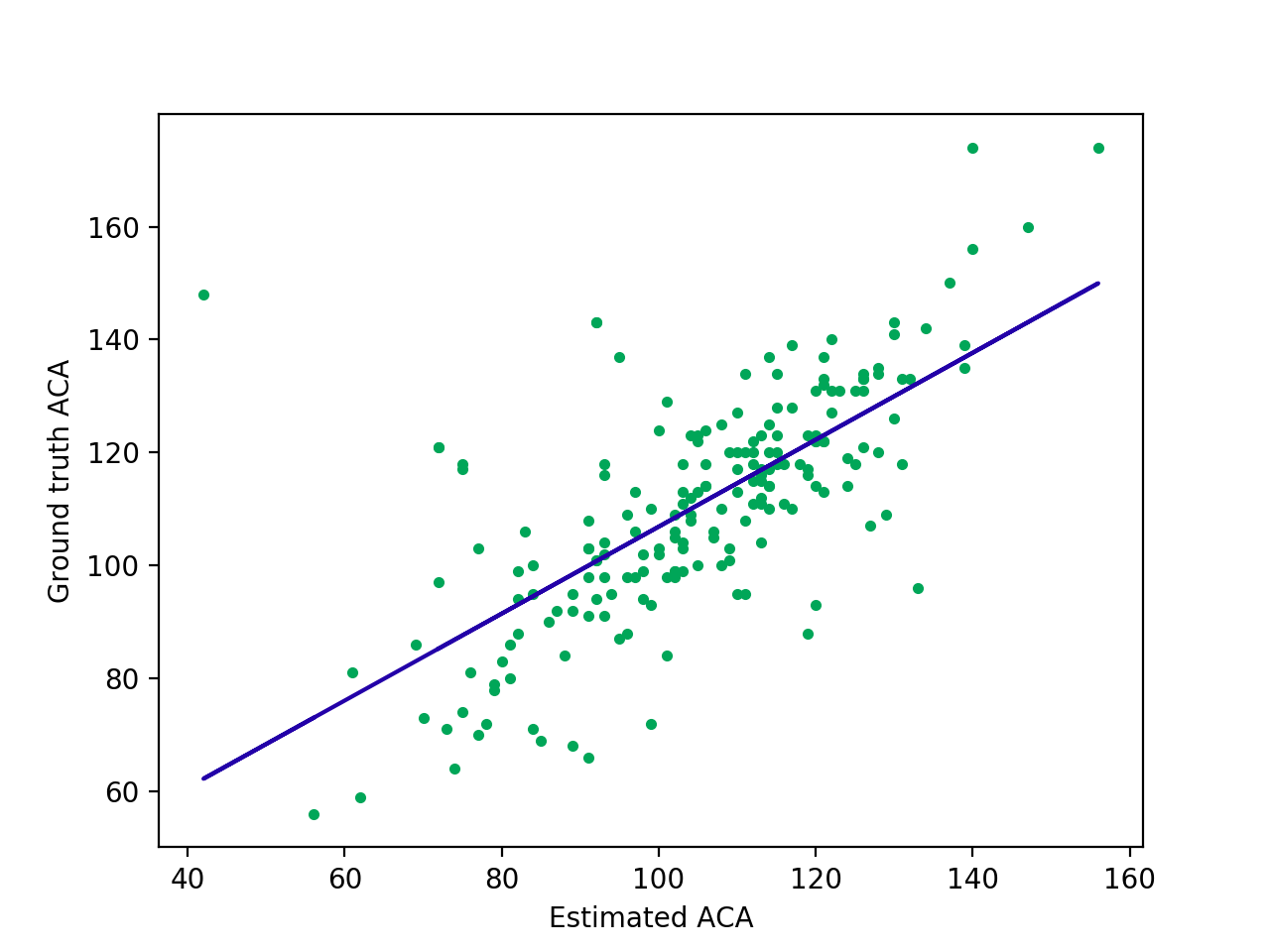}}
  \centerline{(b) ACA estimation}\medskip
\end{minipage}

\caption{Evaluation of our estimation of: (a) cervical length (CL) and (b) anterior cervical angle (ACA). We obtain a RMSE of 110.88, correlation of 0.94 for cervical length and RMSE of 16.22 and correlation of 0.693 for the anterior cervical angle.}
\label{fig:cl_estimation}

\end{figure}

\begin{figure}[ht!]
\begin{minipage}[b]{.28\linewidth}
  \centering
  \centerline{\includegraphics[width=1.25\linewidth]{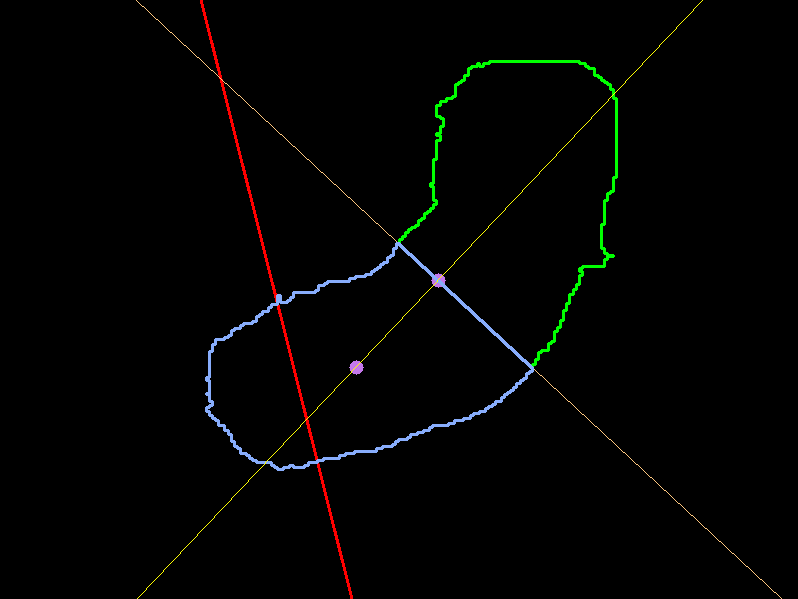}}
  \centerline{1st iteration}\medskip
\end{minipage}
\hfill
\begin{minipage}[b]{0.28\linewidth}
  \centering
  \centerline{\includegraphics[width=1.25\linewidth]{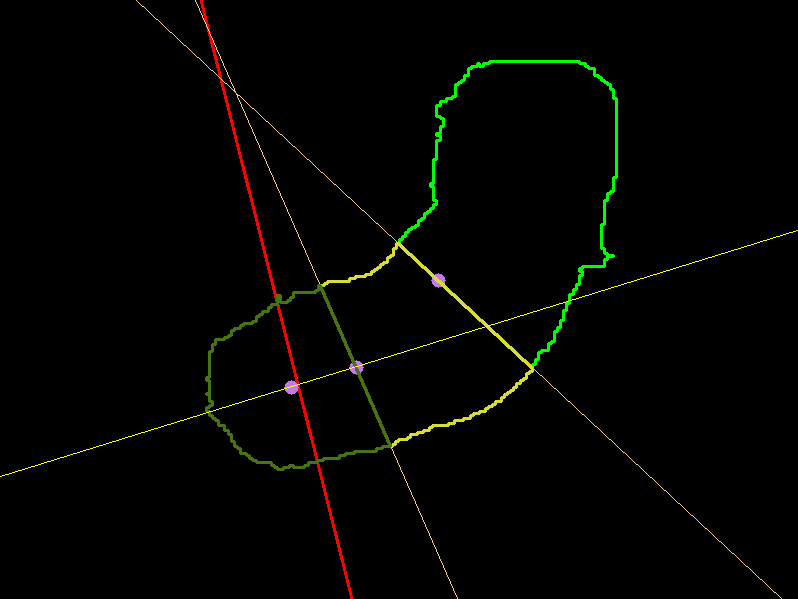}}
  \centerline{2nd iteration}\medskip
\end{minipage}
\hfill
\begin{minipage}[b]{0.28\linewidth}
  \centering
  \centerline{\includegraphics[width=1.25\linewidth]{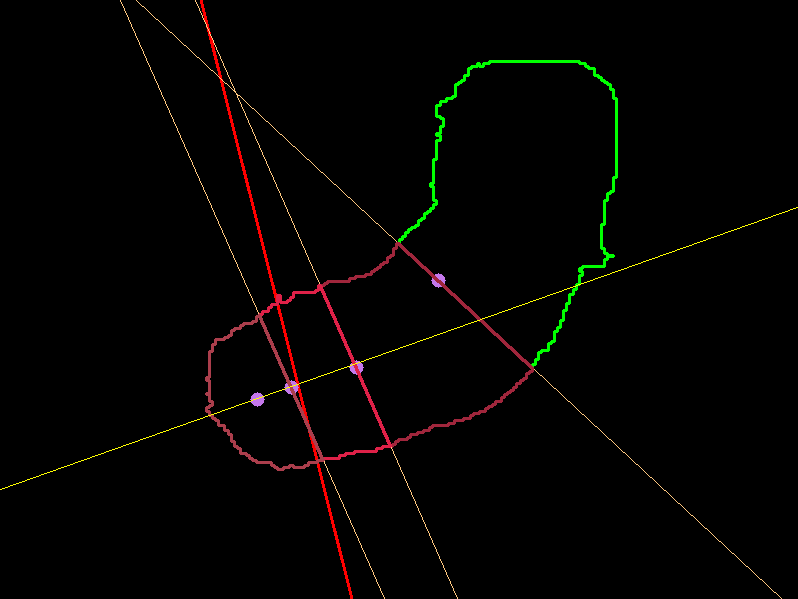}}
  \centerline{3rd iteration}\medskip
\end{minipage}

\caption{Three first iterations of ACA estimation algorithm. On every iteration algorithm finds centroid point of a given shape, splits it into two shapes and proceeds further with the same steps. In the end, we measure ACA between the anterior wall (red line) and the line between the last two centroids.}
\label{fig:aca_iter}
\end{figure}

\textbf{Cervical length estimation:} For this task we use obtained cervix segmentation masks and perform centerline algorithm \cite{C17} on that image set. Then we evaluate whether the cervical length can be estimated by centerline length by conducting linear regression between estimated and ground truth lengths of cervix. We obtain a RMSE of 110.88 and a correlation coefficient of 0.94 what show that these two sets are almost linearly dependent with constant offset. The results are presented in Fig. \ref{fig:cl_estimation}a.
\newline
\hfill \break
\textbf{Anterior cervical angle estimation:} For this task we develop an algorithm which we apply to binary segmentation mask in order to obtain an estimation of Anterior Cervical Angle. Such algorithm is a recursion where on each step we split obtained cervical mask in two parts, based on its centroid location. We perform three iterations of that algorithm on every binary mask. Fig. \ref{fig:aca_iter} presents results of each iteration. Then we evaluate whether our approach can be used to estimate anterior cervical angle by conducting linear regression between estimated and ground truth dataset. We obtain a RMSE of 16.22 and a correlation coefficient of 0.693. The results are presented in Fig. \ref{fig:cl_estimation}b.

\subsection{Preterm birth prediction}

In this section we evaluate classification algorithm on cervical lengths and anterior cervical angles, to assign preterm \textit{vs.} control label to the (CL, ACA) pair. For this purpose, we used four popular machine learning algorithms: Support Vector Machines (SVM), K-Nearest Neighbour, Naive Bayes and Decision Trees. We used the above algorithms for classification due to the fact that they perform well with this type of data.

The best results in terms of accuracy were obtained for classifiers, which were trained on data containing CL and ACA features of the first and second trimesters. This is due to the greater number of features in the set, thus increasing the diversity, which allows for better separation of classes in binary classification.

Despite the simplicity of the naive Bayes classifier, surprisingly high results were obtained, both by analyzing the measures of accuracy, precision and sensitivity for both classes. In addition we conducted a 5-fold cross validation and we obtained the result of accuracy 0.77, confirming the superiority of Bayes classifier. Using this classifier, the highest probability was also obtained that the classifier would determine a randomly chosen positive example higher than the randomly selected negative example, based on the AUC score. Perhaps using the naive Bayes classifier the best results were obtained due to the small correlations between features. 

The worst average results in terms of accuracy, precision, sensitivity and area under the ROC curve were obtained by using the algorithm K-nearest neighbors for classification. The probable reason is the small distance of the samples from each other, which significantly reduces the efficiency of the algorithm.

At this stage, to overcome fact that our ultrasound dataset, after balancing procedure, is very small and it could be a vital reason for poor performance of mentioned four algorithms, we decide to use a different dataset. It contains 380 balanced numerical samples with precomputed cervical length and anterior cervical angle for first and second trimester. It was obtained from King’s College Hospital and Warsaw Medical University.

According to paper \cite{C16}, we have got better results in the classification of spontaneous birth preterm than can be done manually by gynecologists. For the first trimester, we obtained 18\% of false negatives, where manually it is 30\%. This, in turn, can lead to significant time savings and increase the efficiency of prevention treatment.

\renewcommand\arraystretch{1.2}
\newcolumntype{M}{>{\centering\arraybackslash}m{2cm}}
\newcolumntype{C}{>{\centering\arraybackslash}p{1.6cm}}

\begin{table}[ht]
\centering
\caption{Classification results for four different classifiers}
\begin{tabular}{MCCCCC}
\toprule
classifier & trimester & accuracy & precision & recall & AUC\\ 
\midrule
\multirow{3}{*}{SVM}
 & I & 69.56 & 77.0 & 65.0 & 70.19\\
 & II & 62.28 & 65.0 & 68.0 & 61.75\\
 & I + II & 72.5 & 71.0 & 75.0 & 72.5\\
\hline
\multirow{3}{*}{KNN}
& I & 71.74 & 78.0 & 73.0 & 72.1\\
& II & 58.77 & 61.0 & 69.0 & 57.75\\
& I + II & 72.5 & 75.0 & 78.0 & 71.43\\
\hline
\multirow{3}{*}{Naive Bayes}
 & I & 73.91 & 82.0 & 69.0 & 74.62\\
 & II & 59.64 & 61.0 & 73.0 & 58.4\\
 & I + II & \textbf{77.5} & 85.0 & 74.0 & \textbf{78.13}\\
\hline
\multirow{3}{*}{Decision Trees}
 & I & 69.56 & 83.0 & 58.0 & 71.34\\
 & II & 59.65 & 61.0 & 69.0 & 58.72\\
 & I + II & 75.0 & 88.0 & 65.0 & 78.13\\
\bottomrule
\end{tabular}
\end{table}%

\begin{table}[b]
\centering
\caption{Confusion matrix}
\noindent
\renewcommand\arraystretch{1.5}
\setlength\tabcolsep{2pt}
\begin{tabular}{cc|cc}
\multicolumn{2}{c}{}
            &   \multicolumn{2}{c}{Predicted} \\
    &       &   Control &   Preterm             \\ 
    \cline{2-4}
\multirow{2}{*}{\rotatebox[origin=c]{90}{Actual}}
    & Control   & 46   & 16                 \\
    & Preterm    & 21    & 31                \\ 
    \cline{2-4}
    \end{tabular}
\label{tab:cm}
\end{table}

In Table~\ref{tab:cm}, we presented the confusion matrix after classification on numerical data using the naive Bayesian classifier algorithm. We obtained 18\% of false negatives and 14\% false positives for the best classification results. \newline The false negative ratio in our study is higher than the one in \cite{C11}, since we balanced our dataset (it was unbalanced in \cite{C11} which leads to the accuracy paradox and precision and recall bias.) Still, our reported detection rate is 74\% - much higher than 54.8\% reported in \cite{C11}.

\section{Conclusions}
In this paper we propose a method to automatically extract and estimate two biophysical ultrasound markers: CL and ACA based on usage of convolutional neural network. In addition we show that those markers combined can be promising predictor of preterm birth. 
The results presented in this paper show that methods based on deep neural networks can provide automatic, quantitative analysis of ultrasound images. This, in turn, can lead to significant time savings and increase the efficiency of current diagnostic methods without losing its precision.

As future work, we plan to focus on predicting preterm birth with different biophysical markers like shape of cervix or cervix tissue density and on preparing end-to-end method for segmentation and classification task as well.

%
%
%
%

\end{document}